\documentclass[journal]{IEEEtran}
\ifCLASSINFOpdf
  \usepackage[pdftex]{graphicx}
  \DeclareGraphicsExtensions{.pdf,.jpeg,.png}
\else
\fi
\usepackage{amsmath}
\usepackage{url}

\usepackage{xcolor}

\usepackage{cite}

\begin{document}
%
\title{Using Cell Phone Pictures of Sheet Music To Retrieve MIDI Passages}
%
%
%

\author{TJ~Tsai,~\IEEEmembership{Member,~IEEE,}
        Daniel~Yang,~Mengyi~Shan,~Thitaree~Tanprasert,~Teerapat~Jenrungrot
\thanks{Citation information: DOI 10.1109/TMM.2020.2973831, IEEE Transactions on Multimedia.
	
(c) 2020 IEEE.  Personal use of this material is permitted.  Permission from IEEE must be obtained for all other uses, in any current or future media, including reprinting/republishing this material for advertising or promotional purposes, creating new collective works, for resale or redistribution to servers or lists, or reuse of any copyrighted component of this work in other works.

T. Tsai is with the Department
of Engineering, Harvey Mudd College, Claremont,
CA, 91711 USA e-mail: ttsai@hmc.edu.}
}

%
%

\markboth{Journal of \LaTeX\ Class Files,~Vol.~14, No.~8, August~2015}%
{Shell \MakeLowercase{\textit{et al.}}: Bare Demo of IEEEtran.cls for IEEE Journals}
%



\maketitle

\begin{abstract}
This article investigates a cross-modal retrieval problem in which a user would like to retrieve a passage of music from a MIDI file by taking a cell phone picture of several lines of sheet music.  This problem is challenging for two reasons: it has a significant runtime constraint since it is a user-facing application, and there is very little relevant training data containing cell phone images of sheet music.  To solve this problem, we introduce a novel feature representation called a bootleg score which encodes the position of noteheads relative to staff lines in sheet music.  The MIDI representation can be converted into a bootleg score using deterministic rules of Western musical notation, and the sheet music image can be converted into a bootleg score using classical computer vision techniques for detecting simple geometrical shapes.  Once the MIDI and cell phone image have been converted into bootleg scores, we can estimate the alignment using dynamic programming.  The most notable characteristic of our system is that it has no trainable weights at all --- only a set of about 40 hyperparameters.  With a training set of just 400 images, we show that our system generalizes well to a much larger set of 1600 test images from 160 unseen musical scores.  Our system achieves a test F measure score of $0.89$, has an average runtime of $0.90$ seconds, and outperforms baseline systems based on music object detection and sheet--audio alignment.  We provide extensive experimental validation and analysis of our system.
\end{abstract}


%
\IEEEpeerreviewmaketitle

\section{Introduction}
%
%
%
%
\IEEEPARstart{C}{ross}-modal retrieval has been an active area of research in the multimedia community in recent years.  A large number of works have focused on retrieval between images and text (e.g. see \cite{hu2019deep, song2019deep, yu2019adaptive, zhang2018generalized, hu2018twitter100k} for a sample of recent work), facilitated by the availability of large-scale datasets such as Wiki \cite{rasiwasia2010new}, MIRFlickr \cite{huiskes08}, NUS-WIDE \cite{chua2009nus}, Microsoft COCO \cite{lin2014microsoft}, and Pascal Sentences \cite{rashtchian2010collecting}.  Many other works have explored novel cross-modal retrieval scenarios, such as face retrieval across image and video domains \cite{jing2018heterogeneous}, finding multimedia objects with geo-tags using distance proximity and semantic concept similarity \cite{zhu2019efficient}, matching voices to faces \cite{nagrani2018learnable}, using ingredients and cuisine information to retrieve food images \cite{min2016being}, using handdrawn sketches to retrieve photos \cite{xu2018cross}, and matching audio with music video images \cite{wu2016bridging}.

This article explores a novel multimodal application that allows a person to retrieve a passage of music in a natural and expressive way.  Imagine a person sitting at a piano learning a new piece of music.  She wants to know what a particular passage of music sounds like, so she takes out her cell phone and takes a picture of a few lines of music on the physical page of sheet music sitting in front of her.  She is immediately able to hear what those lines of music sound like, and perhaps also listen to different interpretations of the passage by various musical artists.  We study this problem under two simplifying assumptions: we assume that the piece is known (i.e. the person knows which piece of music they are learning) and that a symbolic representation of the piece such as MIDI is available.\footnote{MIDI stands for musical instrument digital interface.  A MIDI file encodes the onset time, pitch, duration, and volume of all notes.}  This is a cross-modal retrieval problem in which the input is a cell phone image and the result is a temporal segment of the symbolic music representation.  This is the camera-based sheet--MIDI passage retrieval problem.

Beyond the inherent challenges common to all cross-modal retrieval problems, there are several constraints in this problem that make it challenging.  First, there is a significant runtime constraint.  Because this is an online, user-facing application, the time it takes to return a result is a very important factor.  A user will not be willing to wait more than a few seconds from the time the picture is taken until they hear the corresponding passage of music.  In this work, we adopt a maximum 1 second average processing time as a hard constraint.  Second, the feature representation of the image must be very compact.  Because transferring a picture from a cell phone to a server would already take more than 1 second, it is necessary to compute a compact set of features on the mobile device, transfer only the features across the network, and then perform a search on the server.  Given typical upload speeds in 4G cellular networks, this means that the feature representation should be in the tens of kilobytes or less.  Third, there is very little training data available.  While datasets of synthetic sheet music and scanned sheet music exist (e.g. the DeepScores dataset \cite{tuggener2018deepscores}, International Music Score Library Project\footnote{\url{https://imslp.org}, accessed December 2019.}), the resources for cell phone images of sheet music are very limited in size and scope.  Practically speaking, we are limited to the amount of data that we ourselves can collect.  Fourth, annotating cell phone images at anything beyond the level of musical measures is impractical.  In particular, manually annotating images at the level of pixels or bounding boxes is totally infeasible, especially at a scope that would enable one to train a multimodal deep neural network.  An ideal solution, then, should have these characteristics: (a) runtime less than one second, (b) feature representation less than 100KB, (c) require very little training data, and (d) should not require annotations at the level of pixels or bounding boxes.

This article describes a solution to the problem that meets the above criteria.  The key to our approach is a novel feature representation called a bootleg score.  The bootleg score is a sparse, binary symbolic representation that describes the position of musical noteheads relative to staff lines in sheet music.  A symbolic music representation such as MIDI can be converted to a bootleg score using deterministic rules that are dictated by the conventions of Western musical notation.  The cell phone image can be converted to a bootleg score by focusing exclusively on detecting just three types of musical objects: filled noteheads, staff lines, and bar lines.\footnote{Noteheads indicate when and which musical notes should be played, staff lines are sets of horizontal lines that indicate the relative pitch of a note, and bar lines are vertical lines that divide the sheet music into measures.}  Because these three types of objects are all simple geometrical shapes -- circular blobs and straight lines -- they can be detected sufficiently well using classical computer vision techniques.  Once the cell phone image and symbolic music representation have been converted to a bootleg score representation, the matching segment of music can be found using subsequence dynamic time warping (DTW).

The key defining characteristic of our system is that it has no trainable weights at all --- only a set of about 40 hyperparameters.  Nearly all of the operations in our system can be interpreted as layers of a neural network that either have no trainable weights (e.g. min, median, and max pooling operations) or have weights that can be manually designed (e.g. a convolutional layer containing a bank of differently sized comb filters for detecting staff lines).  The hyperparameters, in turn, are all interpretable and can be set very intuitively, and many of them have values that are largely dictated by the conventions of Western sheet music notation.  The main benefit of this approach is that it does not require any annotations at the pixel or bounding box level, and it can be tuned with very little training data.  To demonstrate this benefit, we train our system with just 400 cell phone images and show that the system generalizes well to a much larger set of 1600 test images taken from 160 unseen musical scores.

This article has three main contributions:\footnote{This article is a journal extension of the conference paper \cite{yang2019midipassage}.  Compared to \cite{yang2019midipassage}, we have largely redesigned the extraction of bootleg score features from the cell phone image, which results in nearly an order of magnitude reduction in runtime while simultaneously improving retrieval accuracy.  We have added five additional baselines, including systems based on state-of-the-art music object detection and sheet--audio alignment.  The dataset has also been expanded to twice its original size, and all the analyses in section \ref{sec:analysis} are new.}

\begin{itemize}
	\item \textit{Dataset}.  We collect and annotate a dataset for studying the camera-based sheet--MIDI passage retrieval problem.  This dataset contains 2000 cell phone images, 200 MIDI files, and 200 PDF scores containing scanned sheet music, along with a complete set of measure-level annotations for all three types of data.  The dataset is shared publicly with the research community in the hopes of furthering research on camera-based retrieval of music-related data.\footnote{The dataset and code can be found at \url{https://github.com/tjtsai/SheetMidiRetrieval}.}
	\item \textit{Feature representation}.  We introduce a novel feature representation called a bootleg score which enables cross-modal retrieval between a sheet music image and a symbolic music representation.  The bootleg score representation is extremely compact, it can be computed very efficiently at runtime, it can be adapted to new domains (e.g. scans of sheet music or cell phone pictures of sheet music) with very little training data, and it does not require any annotations at the pixel or bounding box level.  These characteristics make it well-suited to the camera-based sheet--MIDI passage retrieval problem.
	\item \textit{Experimental validation}.  We provide extensive experimental validation of a system based on the bootleg score representation.  We show that this system can be tuned with a very small amount of training data, and that it generalizes well to a much larger set of unseen test images.  The system achieves a test F measure of $0.89$, achieves an average runtime of $0.90$ seconds, and reduces each cell phone image to an average feature size of approximately $800$ bytes.
\end{itemize}

The rest of the paper is organized as follows.  Section 2 discusses the relation to previous work.  Section 3 describes the proposed system in detail.  Section 4 explains the experimental setup.  Section 5 presents the empirical results.  Section 6 provides several analyses of system behavior.  Section 7 concludes the work.

\section{Relation to Previous Work}

In this section, we discuss two areas of related work and explain how the current scenario of interest relates to them.  We focus our discussion on works that are cross-modal and deal with sheet music data.

The first area of related work is optical music recognition (OMR).  OMR has a long history of study, and here we only discuss two recent threads of development that are directly relevant to the current scenario.  One important line of work is the development of large-scale datasets for music object detection, one of the central tasks in an OMR system.  Tuggener et al. \cite{tuggener2018deepscores} introduce DeepScores, a very large dataset of synthetic Western sheet music containing pixel-level annotations for a variety of music symbols.  Hajic and Pecina \cite{hajic2017muscima++} introduce MUSCIMA++, a dataset of handwritten music with over 90,000 manually annotated musical symbols.  The availability of these large-scale datasets has standardized evaluation and enabled researchers to develop deep learning models for music object detection.  Another significant line of work is the exploration of different neural network architectures for music object detection.  Pacha et al. \cite{pacha2018handwritten}\cite{pacha2018optical} present a region-based convolutional neural network that cuts the image into chunks and performs object detection on each chunk using Faster R-CNN \cite{ren2015faster}.  Haji{\v c} et al. \cite{hajic2018towards} perform semantic segmentation with a U-Net architecture \cite{ronneberger2015u} and then apply a subsequent detection stage.  Tuggener et al. \cite{tuggener2018deepwatershed} propose a Deep Watershed Detector, which is a fully convolutional network modified to predict the likelihood, type, and size of objects at each pixel location.  Huang et al. \cite{huang2019state} use the darknet53 basic network in YOLO \cite{redmon2018yolov3} to predict the pitch and duration of notes.  Pacha et al. \cite{pacha2018baseline} characterize the performance of several state-of-the-art large object detection models, including Faster R-CNN \cite{ren2015faster}, RetinaNet \cite{lin2017focal}, and U-Net \cite{ronneberger2015u}.  Going beyond simply detecting objects, a handful of works have also explored framing the OMR system as an end-to-end learning task \cite{shi2016end, van2017optical, calvo2018camera, calvo2019handwritten}, though these works have so far been limited to monophonic melodies.

The second area of related work is sheet--audio and sheet--MIDI alignment.  The goal of this task is to find correspondences between audio/MIDI and sheet music images.  There are three general approaches to the problem.  One approach is to use an existing OMR system to convert the sheet music into a symbolic (MIDI-like) representation, to collapse the pitch information across octaves to get a chroma representation, and then to compare this representation to chroma features extracted from the audio.  This approach has been applied to synchronizing audio and sheet music \cite{KurthMFCC07_AutomatedSynchronization_ISMIR, DammFKMC08_MultimodalPresentationofMusic_ICMI, ThomasFMC12_LinkingSheetMusicAudio_DagstuhlFU, izmirli2012bridging, damm2012digital}, identifying audio recordings that correspond to a given sheet music representation \cite{FremereyMKC08_AutomaticMapping_ISMIR}, and finding the audio segment corresponding to a fragment of sheet music \cite{FremereyCME09_SheetMusicID_ISMIR}.  Another approach (to sheet--MIDI alignment) is to convert MIDI to an image in pixel space that resembles the sheet music, and then to perform alignment between columns of pixel values \cite{tanprasert2019midisheet}.  A third approach has been explored in recent years: directly learning the similarity between the two different modalities using deep learning.  Here, the goal is to train a neural network to embed a short segment of sheet music and a short segment of audio into the same feature space, where similarity can be computed directly.  This approach has been explored in the context of online sheet music score following \cite{dorfer2016live}, sheet music retrieval given an audio query \cite{dorfer2016towards, dorfer2017learning, dorfer2018learningAudioSheet}, and offline alignment of sheet music and audio \cite{dorfer2017learning}.  Dorfer et al. \cite{dorfer2018learningToListen} have also recently shown promising results formulating the score following problem as a reinforcement learning game.  See \cite{mueller2019cross} for a recent overview of work in this area.

\begin{figure}
	\includegraphics[width=\columnwidth]{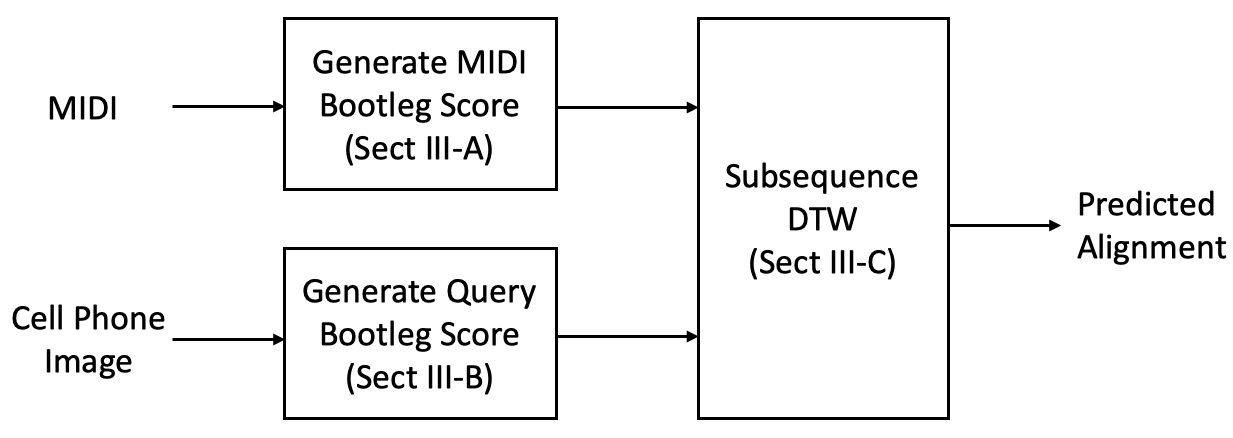}
	\caption{Block diagram of the proposed system.  The main contribution of this work is to introduce a feature representation called a bootleg score which enables cross-modal retrieval between a sheet music image and a symbolic music representation.}
	\label{fig:systemOverview}
\end{figure}

The current scenario of interest differs from these previous works in four significant ways.  First, the current scenario uses \textit{cell phone pictures} of sheet music.  Nearly all of the previous work in OMR and sheet--audio alignment assumes that the sheet music is either a synthetically rendered image or a digital scan of printed sheet music.  Using cell phone images introduces a number of issues that are not present with synthetic sheet music or digital scans: variable lighting conditions, non-perpendicular viewing angle, camera lens distortion and warping, blur, cropping only a portion of the page of sheet music, including background objects outside the page of music, etc.  In recent years, a handful of works have begun to explore OMR on camera-based musical scores \cite{calvo2018camera, bui2014staff, vo2016mrf, blanes2017camera, vo2014distorted}, but this area of study is in its infancy.  Second, the current scenario is an \textit{online} task.  Because OMR is usually framed as an offline task, state-of-the-art OMR systems focus on maximizing recognition accuracy without much regard to the amount of runtime.  For instance, in a recent article that establishes baselines for state-of-the-art music object detection \cite{pacha2018baseline}, the two best performing systems take 40-80 seconds and 20-50 seconds to process each image at inference time \textit{using a GPU}.  These approaches are orders of magnitude too slow to be practical in the current scenario, not even considering the fact that the processing would have to be done on a cell phone.  Nearly all of the above works in sheet--audio alignment are likewise framed as offline tasks.  Third, the current task is a \textit{retrieval} problem.  Whereas the goal in OMR is to correctly interpret \textit{all} of the symbols in sheet music, a retrieval system only needs to extract enough information to uniquely identify a passage.  Given the stringent runtime constraint, it is preferable to extract the minimum amount of information necessary for identification.  Fourth, the current scenario has little or no relevant training data.  The only relevant dataset that we are aware of is Camera-PrIMuS \cite{calvo2018camera}, but this is a synthetic dataset that imitates the appearance of cell phone pictures and it only contains single lines of simple monophonic melodies.  While collecting and annotating cell phone pictures at the measure level is feasible (albeit time-consuming), annotating images at the pixel or bounding box level is totally impractical, especially at the size and scope required to train or fine-tune a multimodal deep neural network.  Due to the lack of labeled training data, state-of-the-art methods for OMR and sheet--audio alignment are not viable solutions to the current problem.  Because of these significant differences between previous work and the current scenario, we opted to design our solution from scratch, keeping runtime performance as a primary consideration.  We will, however, incorporate many of these previous works as baselines in our experimental simulations.

\section{System Description}
\label{sec:system}

Our system takes two inputs: a cell phone picture of several lines of sheet music and a MIDI file of the corresponding piece.  The output of the system is a prediction of the temporal segment in the MIDI file that matches the lines of sheet music shown in the cell phone picture.  Note that in this problem formulation, we assume that the piece is known, and that we are trying to identify the matching passage of music in the piece.  In our study, we focus exclusively on piano music.

Our approach has three main components, which are shown in Figure \ref{fig:systemOverview}.  The first main component is to convert the MIDI file into a representation which we call a bootleg score.  A bootleg score is a very low-dimensional representation of music which is a hybrid between sheet music and MIDI.  It is a manually designed feature space that explicitly encodes the rules of Western musical notation.  The second main component is to convert the cell phone image into a bootleg score.  The MIDI bootleg score and the sheet music bootleg score have strong similarities, but are distinct in ways that will be discussed later.  The third component is to temporally align the two bootleg scores using subsequence DTW.  Each of these three main components will be discussed in more detail in the following three subsections.

\subsection{Generating MIDI Bootleg Score}
\label{sec:genMidiBootleg}

\begin{figure}
	\includegraphics[width=\columnwidth]{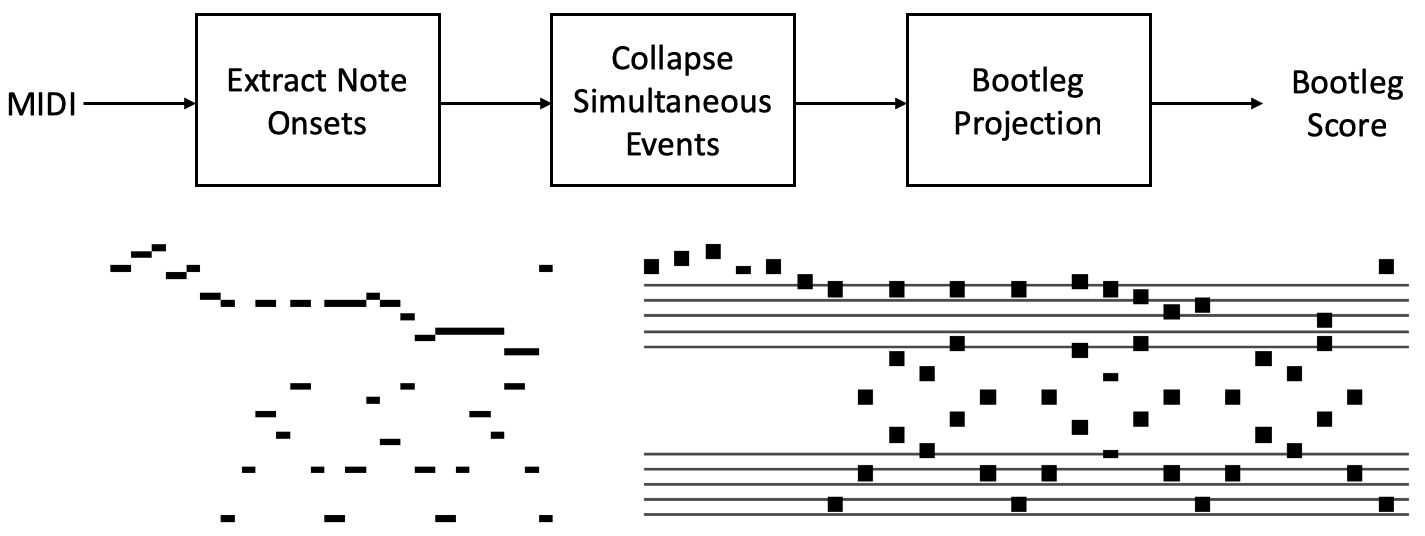}
	\caption{Overview of generating the MIDI bootleg score (section \ref{sec:genMidiBootleg}).  Below the block diagram, a short MIDI passage (left) and its corresponding bootleg score (right) are shown.}
	\label{fig:generateMidiBootleg}
\end{figure}

Generating the MIDI bootleg score consists of the three steps shown in Figure \ref{fig:generateMidiBootleg}.  The first step is to extract a list of all individual note onsets.  The second step is to group the note onsets by their quantized note onset times.  After this second step, we have a list of note events, where each note event consists of one or more (approximately) simultaneous note onsets.  The third step is to project this list of note events into the bootleg feature space.  The bootleg feature representation is a hybrid between sheet music and MIDI, and its similarity to each of the two modalities will be discussed in the next two paragraphs.

The bootleg feature representation can be thought of as a very crude version of sheet music (thus the name ``bootleg score").  It asks the question, ``If I were to look at the sheet music corresponding to this MIDI file, where would the notehead for each note onset appear among the staff lines?"  Note that there is ambiguity when mapping from a MIDI note value to a position in a staff line system.  For example, consider a note onset with note value 60 (C4).  This note could appear in the sheet music as a C natural or a B sharp.\footnote{It could also appear as a D double flat, but we do not consider double sharps or double flats since they occur relatively infrequently.}  Furthermore, in the context of piano music, it could appear in the right hand staff (i.e. one ledger line below a staff with treble clef) or the left hand staff (i.e. one ledger line above a staff with bass clef).  In a similar manner, we can determine a finite set of possible notehead locations for each MIDI note value using the rules and conventions of Western musical notation.  The bootleg feature representation handles this ambiguity by placing a rectangular notehead at all possible locations.  The bootleg score is a binary image representation consisting of these floating rectangular noteheads.

The bootleg feature representation can also be thought of as a symbolic data representation similar to MIDI.  The main drawback of a raw image pixel representation is that it is very high-dimensional, whereas a symbolic format encodes information in a compact way.  The bootleg score captures visual information about the sheet music images, but it encodes that information in a symbolic format.  This allows for a drastic reduction in dimensionality in both the vertical and horizontal dimensions of the bootleg score ``image."  Along the vertical dimension, it represents each staff line location as a single bootleg pixel (which we will refer to as a ``bixel" to differentiate between high-dimensional raw image pixels and low-dimensional bootleg score pixels).  So, for example, two adjacent staff lines in the bootleg score would span a total of three bixels: one bixel for the lower staff line, one bixel for the upper staff line, and one bixel for the position in between.  The bootleg score contains both right hand and left hand staves, where the right hand staff spans from E3 to C8 ($34$ bixels) and the left hand staff spans from A0 to G4 ($28$ bixels).  In total, the bootleg score is 62 bixels tall.  Along the horizontal dimension, we represent each note event as a single bixel column, where a note event consists of one or more simultaneous note occurrences.  We found through experimentation on the training set that a simple modification improves performance in the alignment stage (Section \ref{subsec:alignment}): we simply repeat each bixel column twice and insert an empty bixel column between note events.  This expansion of the data allows the system to recover more easily from local misalignments.

At the end of the three steps shown in Figure \ref{fig:generateMidiBootleg}, we have a bootleg score representation of the MIDI file.  This bootleg feature representation is a $62 \times 3N$ binary matrix, where $N$ is the number of simultaneous note events in the MIDI file.\footnote{The factor of three comes from the filler and repetition.}  The bottom half of Figure \ref{fig:generateMidiBootleg} shows an example of a section of MIDI (left) and its corresponding bootleg score representation (right).  Note that the staff lines in the picture are only included as a visualization aid, but are not present in the actual bootleg feature representation.

\begin{figure}
	\includegraphics[width=\columnwidth]{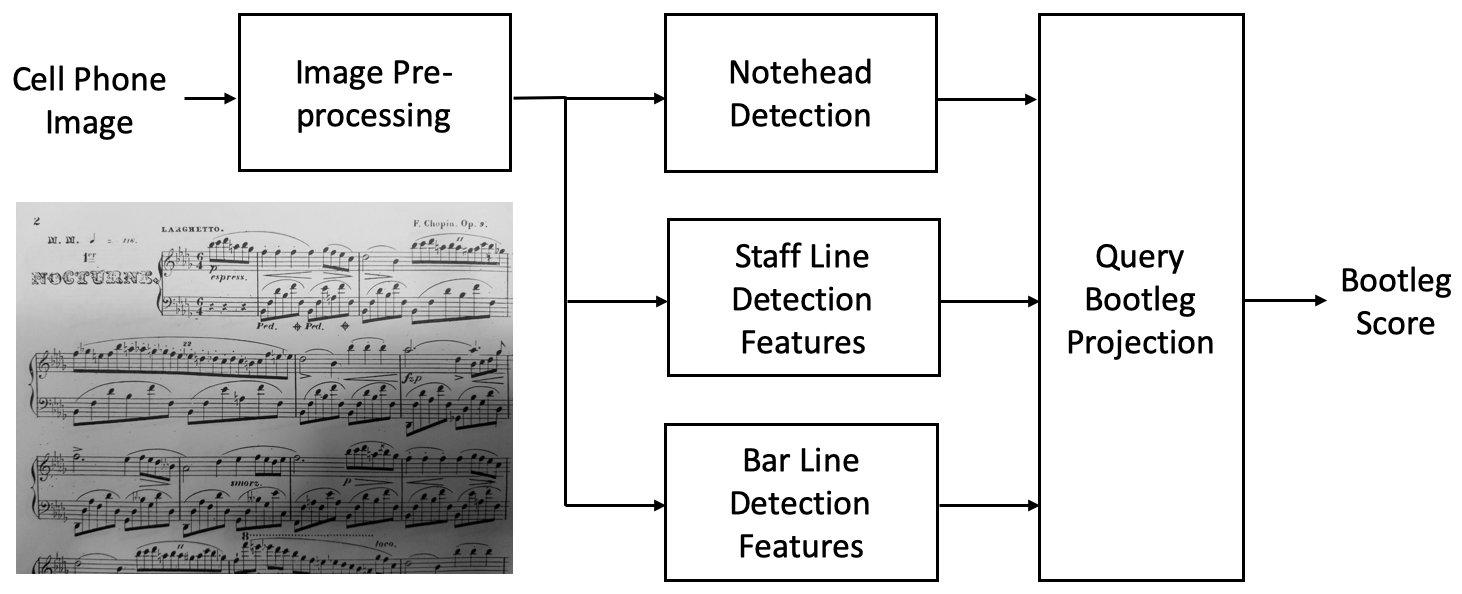}
	\caption{Overview of generating the query bootleg score (section \ref{sec:genQueryBootleg}).  The bottom left shows an example image that will serve as a running example throughout the article.}
	\label{fig:generateQueryBootleg}
\end{figure}

\subsection{Generating Query Bootleg Score}
\label{sec:genQueryBootleg}

The second main component of our system (Figure \ref{fig:systemOverview}) is to convert the cell phone image into a bootleg score representation.  Unlike the MIDI representation, the cell phone image does not explicitly encode any information about the notes.  Therefore, we will have to estimate the notehead information from the raw image.

Our general approach to this task is based on two primary considerations: minimizing runtime and minimizing the need for training data.  The first consideration (minimizing runtime) means that we should detect as little and as simply as possible.\footnote{We note that in the computer vision literature, it is a common technique to speed up processing and retrieval of high-dimensional data by focusing only on selected keypoints (e.g. \cite{lowe1999object}\cite{bay2008speeded}) or keyframes (e.g. \cite{kumar2019evs}\cite{singh2019pics}).}  Our approach is to focus on detecting just three things: filled noteheads, staff lines, and bar lines.  These three types of objects were selected because they are simple geometrical shapes (circular blobs and straight lines) that can be detected reasonably well with classical computer vision techniques.  The second consideration (minimizing the need for training data) means that we should minimize the number of parameters in our system.  In fact, we went so far to the end of the spectrum that our system has \emph{no trainable parameters at all}---it only has a set of approximately $40$ hyperparameters.  This allows us to build a system without doing any pixel-level annotation.

Our method for generating the cell phone image bootleg score has five parts: image pre-processing, notehead detection, staff line detection features, bar line detection features, and bootleg projection.  Figure \ref{fig:generateQueryBootleg} shows how these parts fit together.  Each of these five parts will be described in more detail in the next five subsections.

\subsubsection{Image Pre-processing}
\label{subsubsec:preprocessing}

The preprocessing consists of three operations.  First, we convert the image to grayscale.  Second, we remove background lighting from the image.  We accomplish this by blurring the image and subtracting the blurred image from the non-blurred image.  Third, we resize the image so that the spacing between adjacent staff lines is approximately 10 pixels.  We estimate the staff line separation in four steps: (1) we break the image into a fixed number of columns, (2) we calculate the median pixel value in each row of the columns, (3) we convolve the row medians with a set of comb filters corresponding to different staff line spacings, and (4) we determine the comb filter with the strongest cumulative response from all columns.

\begin{figure}
	\includegraphics[width=\columnwidth]{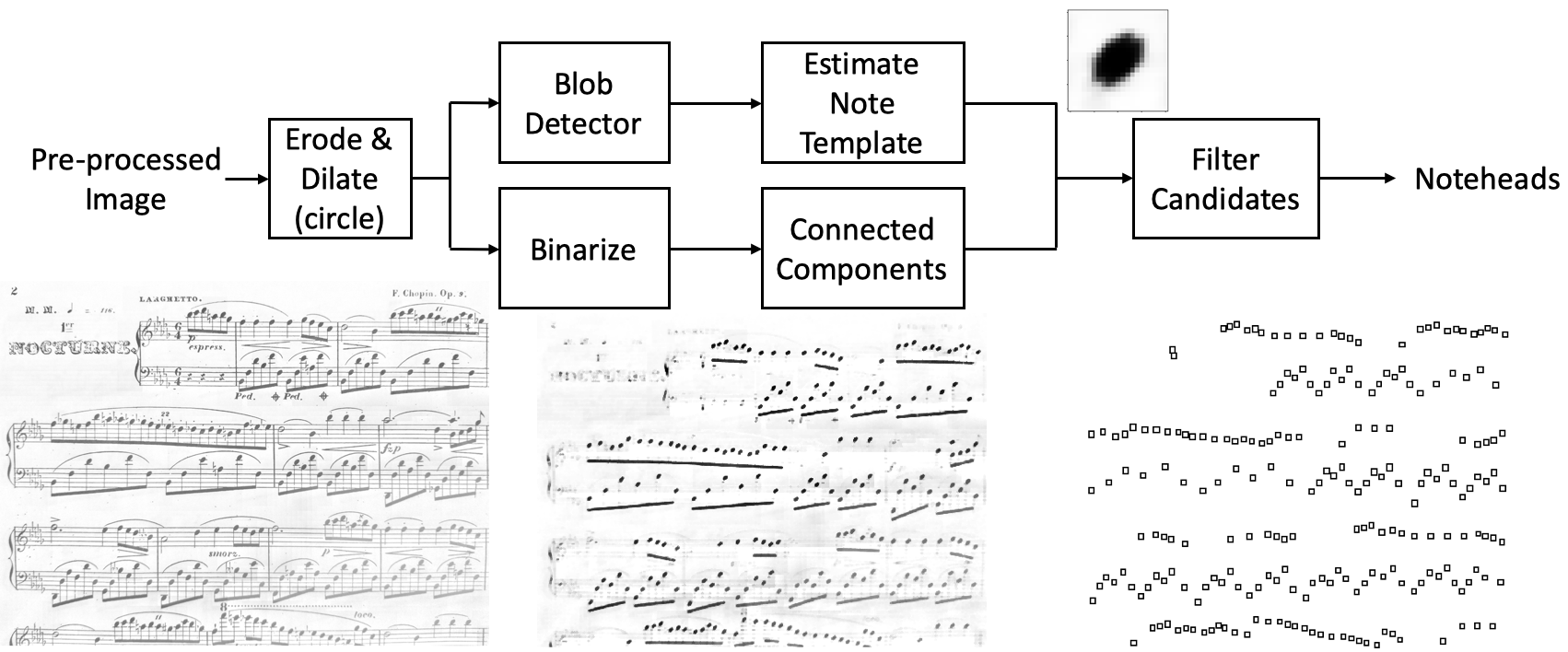}
	\caption{Overview of notehead detection.  The images at bottom show the pre-processed image (left), the result after erosion \& dilation with a circular filter (center), and the detected noteheads (right).}
	\label{fig:noteheadDetection}
\end{figure}

\subsubsection{Notehead Detection}
\label{subsubsec:noteheadDetection}

The goal of the notehead detection stage in Figure \ref{fig:generateQueryBootleg} is to predict a bounding box around every filled notehead in the cell phone image.  We focus on filled noteheads because they are much easier to detect, and they also generally occur much more frequently than half or whole notes.  The notehead detection consists of the steps shown in Figure \ref{fig:noteheadDetection}.  We will explain each of these steps in the next several paragraphs.

The first step is to perform erosion and dilation of the pre-processed image with a circular morphological filter.  The erosion replaces each pixel with the whitest pixel in a circular region centered around the pixel.  This operation removes any objects that consist of thin lines, and it only passes through contiguous dense regions of black pixels.  In practice, we find that there are three types of objects that survive the erosion: filled noteheads, thick note beams (i.e. the bar that connects a sequence of eighth notes) and background pixels (e.g. the music stand outside the boundaries of the page of sheet music).  The dilation takes the resulting image and replaces each pixel with the blackest pixel in a circular region center around the pixel.  This operation restores any objects that survived the erosion back to their original size.  Figure \ref{fig:noteheadDetection} shows an example of an image after erosion and dilation (center image).

Next, we describe the processing in the upper path of Figure \ref{fig:noteheadDetection}.  We take the eroded and dilated image and apply simple blob detection.  We use the simple blob detector in OpenCV with default parameter settings.  The only parameter setting we modify is the minimum and maximum area, which can be set based on the fixed staff line spacing.  We then take crops of the (eroded and dilated) image around the detected keypoints, and we compute the average of the cropped regions.  This average gives us an estimate of what a filled notehead looks like in this image.  Figure \ref{fig:noteheadDetection} shows an example of an estimated template (upper right).

Now we describe the processing in the lower path of Figure \ref{fig:noteheadDetection}.  We take the eroded and dilated image and binarize it using Otsu binarization \cite{otsu1979threshold}.  We then extract a list of connected component regions from the binary image.  At the end of the lower path of Figure \ref{fig:noteheadDetection}, we have a list of candidate regions, some of which are noteheads.

The last step in notehead detection is to filter the list of candidates using our estimated notehead template.  Since our template gives us an estimate of the notehead dimensions, we filter the list of candidates to only contain those regions whose height, width, height-width ratio, and area all roughly match the notehead template (to within some multiplicative tolerance factor).  We found that this approach works very well in identifying isolated notehead blobs, but it fails to identify noteheads in block chords, where multiple noteheads appear in close proximity to one another.  To address this issue, we created a separate set of specifications to identify chord blocks, which often appear as a single connected component region.  Note that the specifications for identifying a block chord can be set very intuitively based on common conventions in piano music.  For example, the minimum area for a block chord is twice the area of the notehead template (with a tolerance factor), the maximum area is five times the area of the notehead template (since a hand generally does not play more than five simultaneous notes at once), the maximum width of a chord block is twice the width of the notehead template, etc.  When a chord block is identified, we would like to determine the bounding box around the individual noteheads in the chord, not just a bounding box around the chord block.  To do this, we simply estimate the number of notes in the chord based on its area relative to the notehead template, and then perform a simple k-means clustering on the individual pixel coordinates in the connected component region.

At the end of these steps, we have a list of bounding boxes around the detected notes in the cell phone image.  Figure \ref{fig:noteheadDetection} shows an example of the predicted notehead locations in an image (lower right).  

\subsubsection{Staff Line Detection Features}

The goal of the staff line detection features stage in Figure \ref{fig:generateQueryBootleg} is to compute a tensor of features that can be used to predict staff line locations in the bootleg projection stage.  Note that unlike the notehead detection, the staff line detection features stage does not actually make a prediction about staff line locations, but it simply constructs the information necessary to make such a decision in the bootleg projection stage.  The reason why we delay this decision to the bootleg projection stage is an important point, and one that we will motivate and explain the next two paragraphs.

Many assumptions that are typically made about staff lines in scanned sheet music do not hold true with cell phone pictures.  We cannot assume that the cell phone picture is taken at a $90$ degree angle to the plane of sheet music, as is the case with digital scans.  We cannot assume that the sheet music is a flat plane, since the physical page of music might itself have some curvature.  The camera lens may also introduce warping at the edges of the picture.  So, the staff lines may not be straight lines in the image, and the staff line spacing may not be constant throughout the image.  If the picture is taken at an angle, for instance, the staff line spacing on the left and right sides of the image may be different.  All of these factors make the staff line detection problem more challenging with cell phone pictures than with digital scans.

Because of the issues discussed above, we estimate staff line locations locally rather than globally.  In other words, for every detected notehead in the cell phone image, we make a local estimate of the staff line location and spacing in the context region around the notehead.  The reason why we delay the staff line location prediction until the bootleg projection stage, then, is because we want to make local predictions that are notehead-centric, rather than global predictions.  This approach addresses the issues discussed above by only assuming that the staff lines behave like straight lines locally rather than globally.

\begin{figure}
	\centerline{\includegraphics[width=\columnwidth]{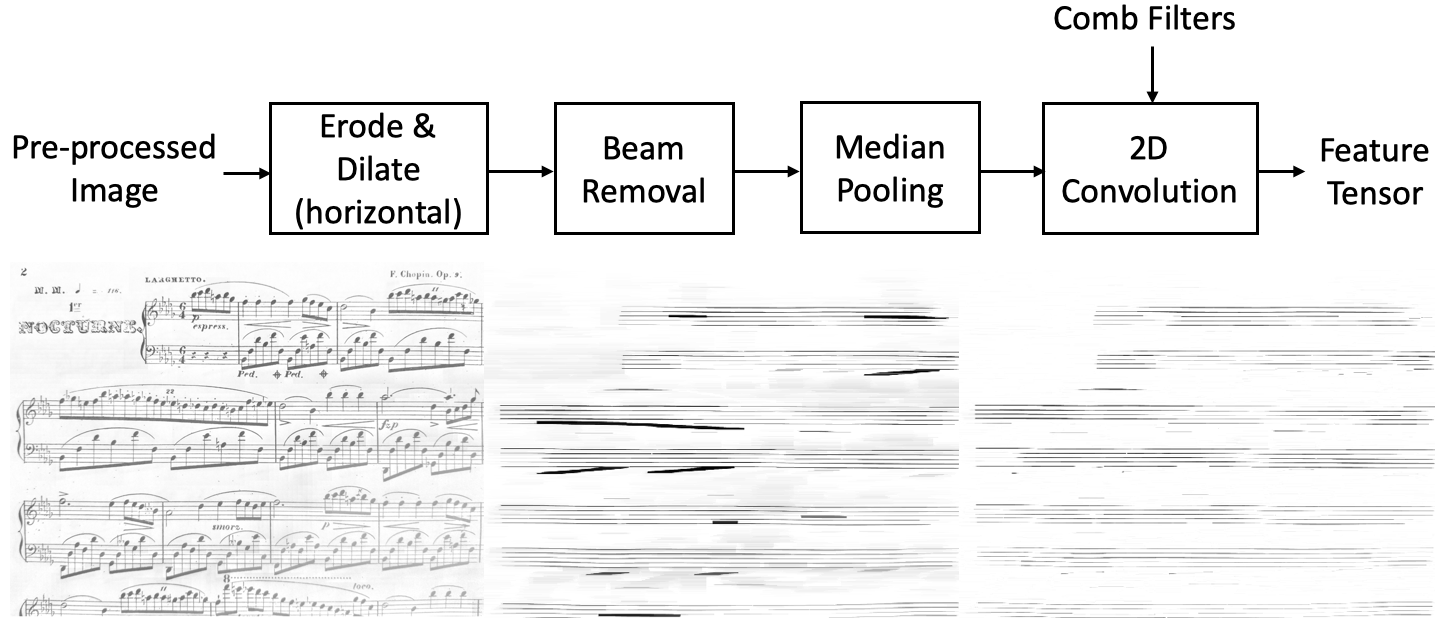}}
	\caption{Overview of staff line features computation.  The images at bottom show the pre-processed image (left), the result after erosion \& dilation with a horizontal filter (center), and the result after removing note beams (right).  The feature tensor is computed by breaking the image into columns, performing median pooling on the column rows, and then convolving with a set of comb filters of various sizes.}
	\label{fig:staffDetectBlock}
\end{figure}

The staff line detection features are computed in four steps as shown in Figure \ref{fig:staffDetectBlock}.  The first step is to perform erosion and dilation on the image with a short ($1$ pixel tall), fat morphological filter.  The purpose of this operation is to filter out everything except for horizontal lines.  In practice, we find that there are two types of objects that survive this operation: staff lines and horizontal note beams (i.e. the beam connecting a sequence of eighth notes).  The second step is to remove the horizontal note beams, as they can throw off the staff line location estimates.  Because the note beams are much thicker than staff lines, we can isolate the note beams based on their thickness (again through morphological filtering) and subtract them away.  The third step is to break the image into a fixed number of columns and to calculate the median pixel value in the rows of the columns.  The fourth step is to convolve the resulting row medians with a set of comb filters corresponding to different staff line spacings.  Since we have already performed interline normalization on the image, the range of comb filter sizes is much narrower than in the image pre-processing stage (section \ref{subsubsec:preprocessing}).  After the convolution step, we have a feature tensor of dimension $K \times H \times C$, where $K$ is the number of comb filters, $H$ is the height of the image in pixels, and $C$ is the number of columns.  This feature tensor contains the activations of a set of differently sized comb filters.

\subsubsection{Bar Line Detection Features}
\label{subsec:barlineDetect}

The goal of the bar line detection features stage (Figure \ref{fig:generateQueryBootleg}) is to calculate features indicating the presence of bar lines.  The bar line locations are needed to correctly cluster noteheads into right and left hand staves in a grand staff.  Similar to the staff line features stage, we delay making a hard decision until the bootleg projection step.

\begin{figure}
	\centerline{\includegraphics[width=\columnwidth]{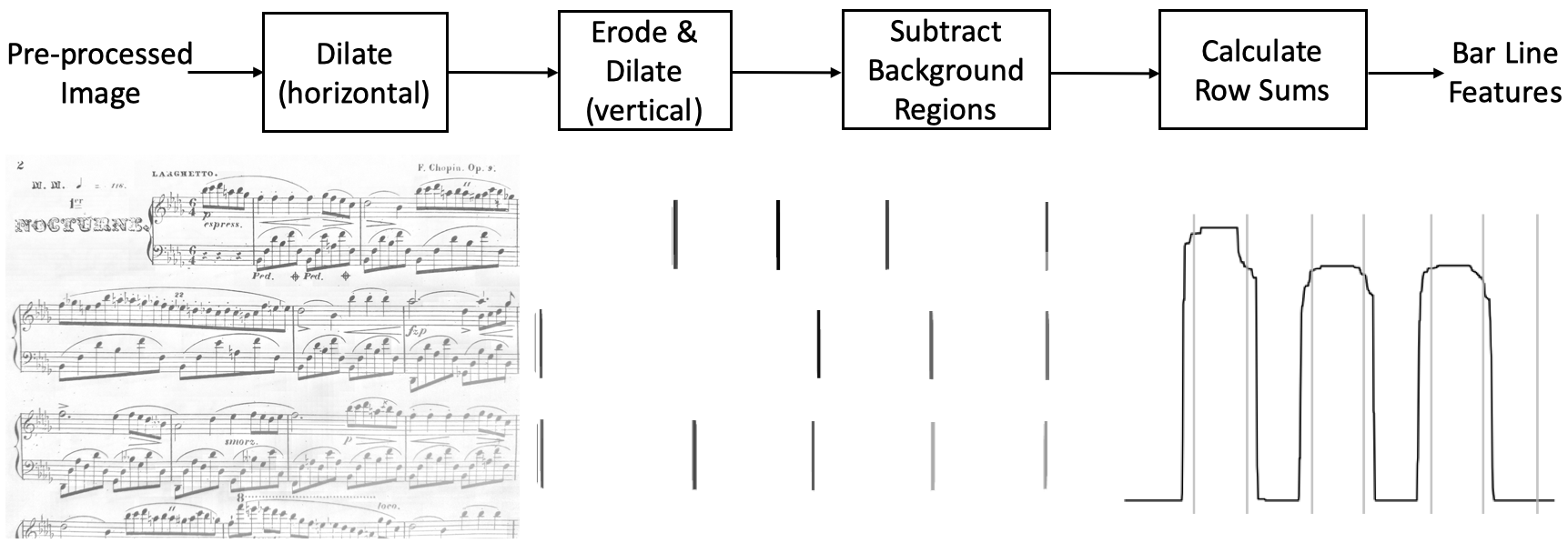}}
	\caption{Overview of bar line detection.  The images at bottom show the pre-processed image (left), the result after isolating the bar lines (center), and the bar line features computed as the row sums (right).}
	\label{fig:barDetectBlock}
\end{figure}

The bar line detection features are computed in four steps as shown in Figure \ref{fig:barDetectBlock}.  The first three steps are designed to isolate bar lines in the image.  The first step performs dilation in the horizontal direction, and the second step performs erosion and dilation with a tall, skinny (1 pixel wide) morphological filter.  The purpose of the first step is to expand bar lines that are warped or slightly angled from vertical, so that they survive the vertical erosion and dilation.  In practice, we find that there are three types of objects that survive these operations: bar lines, very long note stems, and background pixels (e.g. music stand at edges of image).  To remove regions of background pixels, we introduce a third step which subtracts away components that are too thick (which are identified by performing erosion in the horizontal direction).  The fourth step is to calculate the row sum of the resulting image.  The row sums indicate the amount of evidence that a grand staff exists at that row location.  The bottom images in Figure \ref{fig:barDetectBlock} show the isolated bar lines (center) and the corresponding bar line features (right) for an example image.

\subsubsection{Query Bootleg Projection}

\begin{figure}
	\centerline{\includegraphics[width=\columnwidth]{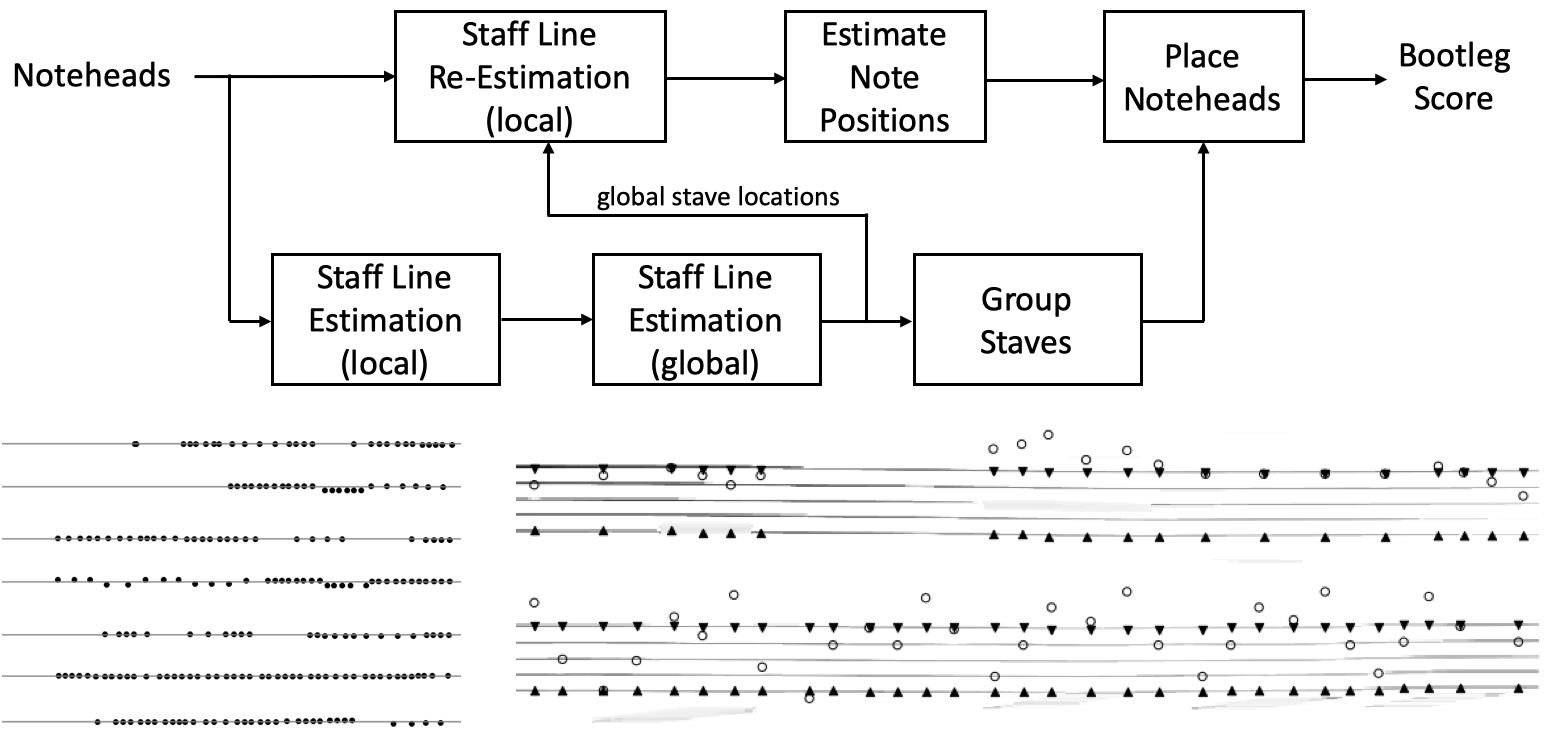}}
	\caption{Overview of query bootleg projection.  The bottom left image shows a global view of the local staff location estimates around each detected notehead in the image (black dots), along with the estimated global staff locations from k-means clustering (gray lines).  These global staff locations are also shown in the bottom right image of Figure \ref{fig:barDetectBlock} as vertical gray lines, where they are used to predict stave grouping.  The bottom right image shows a zoomed in view of the local staff line estimates in a single grand staff.  Each circle corresponds to a detected notehead, and the arrowheads correspond to the estimated staff boundaries.}
	\label{fig:queryBootlegProjection}
\end{figure}

The last step in Figure \ref{fig:generateQueryBootleg} is to combine the notehead, staff line, and bar line information in order to synthesize a bootleg score for the cell phone image.  Figure \ref{fig:queryBootlegProjection} shows the steps involved in synthesizing the bootleg score.  We will discuss these steps in two parts: the staff estimation (three leftmost blocks in Figure \ref{fig:queryBootlegProjection}) and the remaining blocks that focus on integrating information into the actual bootleg score (three rightmost blocks).

The three leftmost blocks in Figure \ref{fig:queryBootlegProjection} estimate the staff locations in two different ways.  The first way is to estimate the staff location and spacing \textit{locally} around each detected notehead.  To do this, we extract a context region around the notehead location in the staff line feature tensor, and then identify the maximum element in this selected context region.  The location of the maximum element specifies both the vertical location of the staff system (i.e. \textit{where} the comb filter had the strongest response), along with the staff line spacing (i.e. \textit{which} comb filter had the strongest response).  One problem with this method of estimation is that many noteheads (particularly those on ledger lines) have two valid staff line systems in their context region and may be incorrectly assigned to a neighboring staff line system.  This problem motivates the need for two different estimation methods.  The second way is to roughly estimate the staff locations \textit{globally}.  The local staff location estimates provide a set of scalar row coordinates that are highly clustered around the staff line locations.  Therefore, we perform a simple k-means clustering, and increase the number of clusters until the minimum distance between two centroids falls below a threshold.  This provides us with an estimate of the number of staves in the image, as well as a rough estimate of their vertical location in the image.  Once we have this global estimate of staff locations, we re-estimate the local staff system around each notehead, where we first assign the notehead to its closest global staff system and use a much smaller context region centered around the assigned global staff location.  This largely mitigates the problem of assigning noteheads to the wrong staff system.  Figure \ref{fig:queryBootlegProjection} shows both a global and local view of staff location estimates in an example image, including the estimated k-means centroids (gray horizontal lines in bottom left image).  

The three rightmost blocks in Figure \ref{fig:queryBootlegProjection} integrate all of the estimated information in order to synthesize the actual bootleg score.  The first block (``Estimate Note Position") uses the predicted staff location and spacing around each notehead to estimate the notehead's vertical staffline position.  This can be done using simple linear interpolation.  The second block (``Group Staves") groups the global staff locations (i.e. the centroids estimated from the k-means clustering) into pairs, corresponding to the left and right hand staff systems in a single grand staff.  Given a set of global staff locations $a_0, a_1, \cdots, a_k$, there are two ways to group them into pairs (i.e. $\{(a_0,a_1), (a_2,a_3), \cdots\}$ or $\{(a_1,a_2), (a_3,a_4), \cdots\}$).  To determine which grouping to adopt, we determine which interpretation is more consistent with the bar line information.  Since the left and right hand staff systems in a grand staff will be connected with bar lines, we expect the bar line features in the region between paired staff systems to be higher.  For example, in the bottom right image of Figure \ref{fig:barDetectBlock}, $a_0, a_1, \cdots, a_6$ correspond to the locations of the seven vertical gray lines, and the bar line features are much higher for the $\{(a_0,a_1), (a_2,a_3), (a_4, a_5)\}$ pairings than the $\{(a_1,a_2), (a_3,a_4), (a_5, a_6)\}$ pairings.  Therefore, we compute the intra-pair median bar line feature value for both interpretations and select the interpretation with the higher value.  The last block (``Place Noteheads") constructs the bootleg score.  We collapse the noteheads within each grand staff into a sequence of simultaneous note events, where noteheads are grouped together if the notehead bounding boxes have any horizontal overlap.  We then construct the bootleg score as a sequence of simultaneous note events.

\subsection{Subsequence DTW}
\label{subsec:alignment}

\begin{figure}
	\centerline{\includegraphics[width=\columnwidth]{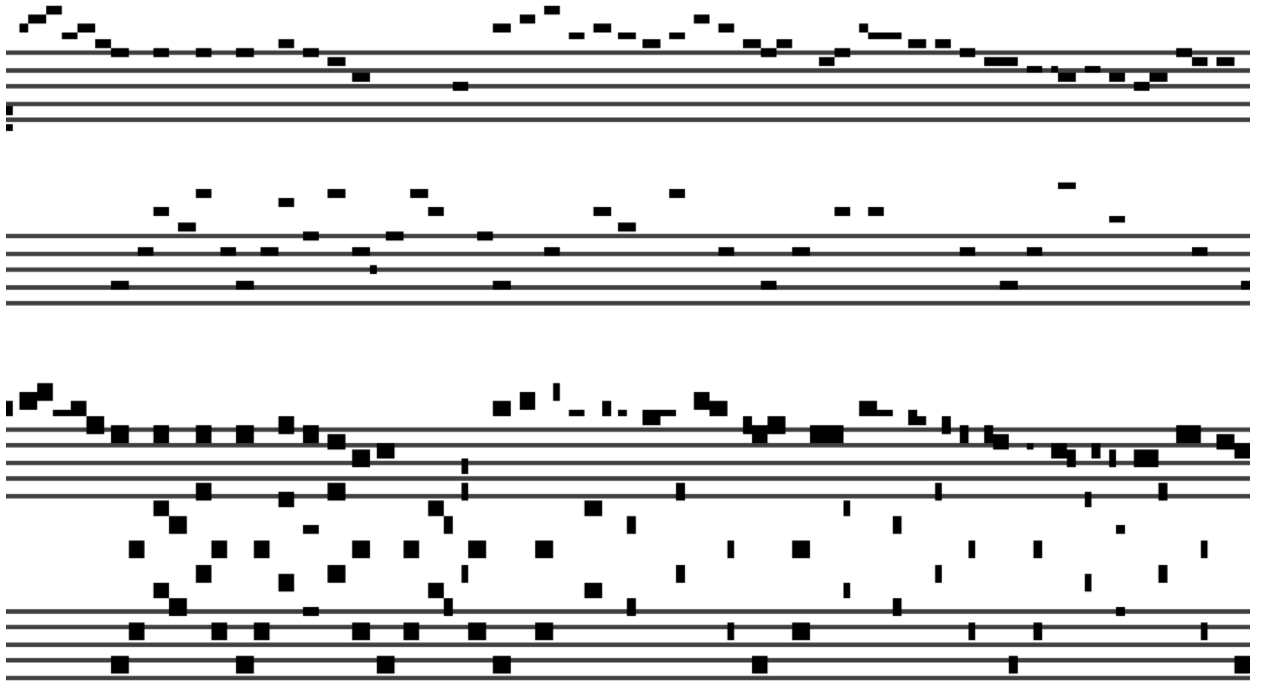}}
	\caption{Sample alignment between a query bootleg score (top) and MIDI bootleg score (bottom).  The staff lines are shown for reference, but are not present in the actual bootleg score representation.  Note that the MIDI bootleg score has many more black elements because it places noteheads at all possible notehead locations.}
	\label{fig:sampleAlignment}
\end{figure}

The third main component of our system (Figure \ref{fig:systemOverview}) is to temporally align the two bootleg scores using subsequence DTW.  DTW is a well-established dynamic programming technique for determining the alignment between two feature sequences.  Subsequence DTW is a variant of DTW that finds the optimal alignment between a short query sequence and a subsequence of a longer reference sequence.  For details on DTW and its variants, the reader is referred to \cite{muller2015fundamentals}.  Our cost metric computes the negative inner product between two bixel columns and then normalizes the result by the maximum of (a) the number of simultaneous noteheads in the sheet music and (b) the number of simultaneous note onsets in the MIDI.  This cost metric ensures that the actual cost is not biased by the number of simultaneous notes.  At the end of this stage, we have a prediction of the segment in the MIDI file that best matches the lines of sheet music shown in the cell phone image.  This is the final output of our proposed system.  Figure \ref{fig:sampleAlignment} shows the alignment between the query bootleg score (top) and MIDI bootleg score (bottom) for the running example.

\subsection{Improvements}
\label{subsec:improvements}	
The system described above incorporates several modifications to \cite{yang2019midipassage} that improve the system's robustness and runtime efficiency.  In the image pre-processing (Figure \ref{fig:generateQueryBootleg}), the adaptive image resizing was introduced to handle different levels of zoom.  In the staff line features computation (Figures \ref{fig:generateQueryBootleg} and \ref{fig:staffDetectBlock}), the median pooling was introduced to drastically reduce the size of the two-dimensional convolution.  The bar line detection approach (Figures \ref{fig:generateQueryBootleg} and \ref{fig:barDetectBlock}) was totally re-designed to improve robustness.  Among other changes, it returns a set of soft features indicating the presence of bar lines, rather than making a hard decision about bar line locations.  The query bootleg projection stage (Figures \ref{fig:generateQueryBootleg} and \ref{fig:queryBootlegProjection}) was also completely re-designed, using both local and global information to improve its estimates.  The process of generating the MIDI bootleg score and the subsequence DTW stage are unchanged (Figures \ref{fig:systemOverview} and \ref{fig:generateMidiBootleg}).

\section{Experimental Setup}

The experimental setup will be described in three parts: the data, the annotations, and the evaluation metric.

The data was collected in the following manner.  We first downloaded $200$ piano scores in PDF format from IMSLP.  These piano scores come from $25$ well-known composers and span a range of eras and genres within the classical piano literature.  To simplify the evaluation, we selected scores that do not have any repeats or structural jumps.   For each score, we then found a corresponding MIDI file from \url{musescore.com} and other various online websites.  This gave us a total of $200$ MIDI-PDF matching pairs.  Next, we printed out the PDF scores onto physical paper, placed the sheet music pages in various locations, and took $10$ cell phone pictures of each score, spaced throughout the length of the piece.  The pictures were taken in various ambient lighting conditions (some of which triggered the flash and some of which didn't), various perspectives, and varying levels of zoom.  The pictures capture between $1$ and $5$ complete lines of music on a page.  We collected the data with four different cell phone cameras (iPhone 8, Galaxy S10, Pixel 3, Moto E4), and all pictures were taken in landscape orientation at an aspect ratio of $4:3$ or $16:9$.  In total, the data consists of $200$ MIDI files, $200$ scores in PDF format, and $2000$ cell phone images.\footnote{Note that, compared to our previous work \cite{yang2019midipassage}, the dataset has been doubled in size.}

The annotations for all three types of data were done at the measure level.  For the MIDI files, we used the Python library \url{pretty_midi} to programatically estimate the timestamps of the downbeats in each measure, which were then manually verified and corrected.  For the (PDF) scores, we annotated the measures that each line of music contains (e.g. the first line of music contains measures 1 through 6, the second line of music contains measures 7 through 11, etc).  And for the cell phone images, we annotated which lines of music in the score were captured. Since the cell phone pictures would often capture a fragment of a line of music at the top or bottom of the image, we adopted the convention of only indicating complete lines of music that are fully captured in the image.  Combining the cell phone image annotations with the score PDF annotations, we can then programatically determine which measure numbers are contained in each image, and therefore which time segment in the MIDI file it corresponds to.

The evaluation metric we use to measure system performance is precision, recall, and F measure.  These are common evaluation metrics for tasks similar to this (e.g. see chapter 4 in \cite{muller2015fundamentals} and \cite{olszewska2019designing}).  Precision is computed by determining the total amount of overlap between the predicted time intervals and ground truth time intervals, and then dividing by the duration of predicted time intervals.  In a handful of instances, there are queries that perfectly match two different sections in the score.  In these situations, we compute the amount of overlap with the ground truth as the maximum overlap between the predicted time interval and \emph{any} of the ground truth time intervals.  To avoid ambiguity, near-identical matching passages that differ by at least one note are considered incorrect.  Recall is computed by determining the total amount of overlap between the predicted time intervals and ground truth time intervals, and then dividing by the total duration of ground truth time intervals.  The F measure is then computed as the harmonic mean of precision and recall.  Note that precision, recall, and F measure all range between $0$ and $1$, where $1$ corresponds to perfect performance.

\section{Results}

We evaluate our system in the following manner.  We first select $40$ out of the $200$ scores and set apart their corresponding $40 \times 10 = 400$ cell phone images as the training set.\footnote{During system development, we discovered that one of the training scores was an arrangement of its corresponding MIDI file rather than an exact match.  This score was discarded, so the final training set contains $390$ images.}  The remaining $1600$ cell phone images are set apart for testing.  Note that this train-test split has an unusually large emphasis on the test data.  The reason that we do this is because our system has no trainable weights, so the training data is really only used for tuning hyperparameters.  After doing iterative design on the training data and determining reasonable hyperparameter settings, we froze the system and evaluated it on the $1600$ test images coming from the $160$ unseen music scores.  The only change that we made during testing was to manually correct the ground truth annotations to account for instances of identical multiple matches.\footnote{We selected the current training set to be a superset of the training set used in our previous work \cite{yang2019midipassage}, in order to ensure that the test data is truly unseen.}

We compare our system to eight different baselines.  These baselines fall into one of four types described below.

The first group of baselines is based on commercial OMR.  We used two different commercial OMR programs (Photoscore and SharpEye) to convert each cell phone image to a predicted MIDI representation, synthesized both the predicted MIDI and the reference MIDI to audio, and then performed subsequence DTW with chroma features to estimate the predicted matching time interval.\footnote{We also experimented with performing the alignment directly on a piano roll representation of the MIDI, but found that synthesizing to audio had better performance.}  Note that Photoscore and SharpEye were not designed to handle cell phone images, so they often produced complete jibberish or simply failed to process the image (i.e. they would throw an error).  In the latter situations, we simply mapped errors to a predicted time interval with $0$ duration.

The second group of baselines is based on state-of-the-art music object detection.   We ran experiments with three different object detectors: Faster R-CNN \cite{ren2015faster}, RetinaNet \cite{lin2017focal}, and Deep Watershed Detector \cite{tuggener2018deepwatershed}.  All three systems were trained on DeepScores \cite{tuggener2018deepscores}, a synthetic dataset of sheet music with pixel-level annotations.  For the Faster R-CNN and RetinaNet models, we use code provided by Pacha et al. \cite{pacha2018baseline} for training state-of-the-art object detection systems on DeepScores.  Note that music object detection by itself does not solve the retrieval task -- it is only one component in a retrieval system.  Accordingly, we use each of these three object detectors in our proposed framework as a plugin replacement for the notehead detection method described in section \ref{subsubsec:noteheadDetection}.\footnote{The original DeepScores dataset does not include staff lines or bar lines among the list of detected objects, so we only use the object detectors to detect noteheads.}  Thus, these baselines can be thought of as variants of our proposed system in which state-of-the-art object detection methods are used to compute the bootleg score representation.  Because the Deep Watershed Detector is the only system that was specifically designed for detecting music symbols, we went to additional lengths to fine-tune this model on a small set of 2420 manually annotated noteheads taken from 40 lines of music in cell phone pictures of 40 different scores.  Thus, there are four baselines in this second group: Faster R-CNN (trained on DeepScores), RetinaNet (trained on DeepScores), Deep Watershed Detector (trained on DeepScores), and fine-tuned Deep Watershed Detector (trained on DeepScores and then fine-tuned on cell phone data).

The seventh baseline is the sheet--audio retrieval system proposed by Dorfer et al. \cite{dorfer2018learningAudioSheet}.  This system first pre-processes a page of sheet music into single lines of music, and then it performs the alignment by projecting chunks of sheet music and chunks of the audio spectrogram into a common embedding space that encodes semantic similarity.  Because this system is trained on synthetic audio data generated from MIDI files, it can be interpreted as a sheet--MIDI retrieval system where the MIDI is first converted to audio as a pre-processing step.  Note that this is the only baseline system that was designed to solve the same cross-modal retrieval problem that we are tackling.

The last baseline is random informed guessing.  We calculate the average number ($N$) of sheet music measures showing in the training images.  At test time we randomly select a time interval in the reference MIDI file spanning $N$ measures.  Because none of the baseline systems have been designed specifically to handle cell phone images, we expect some of the systems to perform very poorly.  This baseline provides a frame of reference to interpret our results by establishing a performance floor.

\begin{table}
	\begin{center}
		\begin{tabular}{| l | l | c | c | c | c |}
			\hline
			System & Data & P & R & F & Runtime\\
			\hline
			Random  & Test & $.155$ & $.162$ & $.158$ & $0.0$s\\
			SharpEye & Test & $.429$ & $.076$ & $.129$ & --\\
			Photoscore & Test & $.643$ & $.624$ & $.633$ & --\\
			RetinaNet \cite{lin2017focal} & Test & $.516$ & $.264$ & $.350$ & $11.7$s\\
			Sheet--Audio Align \cite{dorfer2018learningAudioSheet} & Test & $.687$ & $.284$ & $.402$ & $17.5$s \\
			Faster R-CNN \cite{ren2015faster} & Test & $.837$ & $.870$ & $.854$ & $49.9$s\\
			DWD \cite{tuggener2018deepwatershed} & Test & $.000$ & $.000$ & $.000$ & $221.0$s\\
			DWD (fine-tuned) \cite{tuggener2018deepwatershed} & Test & $.910$ & $.872$ & $.891$ & $213.1$s\\			
			\hline
			Bootleg \cite{yang2019midipassage} & Test & $.874$ & $.815$ & $.843$ & $8.01$s\\
			Bootleg (Improved) & Test & $.894$ & $.885$ & $.889$ & $0.90$s\\	
			Bootleg (Improved) & Train & $.922$ & $.911$ & $.916$ & $0.91$s\\
			\hline
		\end{tabular}
	\end{center}
	\caption{Comparison of system performance.  The four rightmost columns show precision, recall, F measure, and average runtime per query.  The baseline systems have been ordered from top to bottom by average runtime, and the proposed system is indicated at bottom as ``Bootleg (Improved)."}
	\label{tab:results}
\end{table}

Table \ref{tab:results} shows the performance of our system and the eight baseline systems.\footnote{The system described in this article is indicated as ``Bootleg (Improved)."  We also evaluate an older version of the system as originally proposed in \cite{yang2019midipassage} for comparison.  See section \ref{subsec:improvements} for a description of the differences between these two systems.}  There are four things to notice about these results.  First, most baseline systems perform poorly.  This is not a surprise, since the baseline systems were all designed to handle synthetic sheet music or scans of printed sheet music, not cell phone images.  The most extreme example of this is the Deep Watershed Detector, which failed to detect any noteheads on most images and achieved an F measure score of $0$.  The two exceptions to this are Faster R-CNN and the fine-tuned Deep Watershed Detector, which achieve F measure scores of $.854$ and $.891$, respectively.  Second, the proposed system ($.889$ F measure) approximately matches the retrieval accuracy of the best baseline system ($.891$ F measure), but with much less computation.  Third, the proposed system generalizes very well from the training data to the testing data.  After iterating and optimizing the system on the training set, the F measure score only fell from $.916$ (on the training data) to $.889$ (on the test data).  The reason that our system generalizes so well with such a small training data set is that our system has no trainable weights and only about $40$ hyperparameters.  Even then, many of these hyper parameters are dictated by conventions of Western musical notation for piano music.  With such a small number of parameters, we don't expect the system to suffer severely from overfitting, and indeed this is what we observe in our experiments.  Fourth, the proposed system is a considerable improvement over the originally proposed bootleg system in \cite{yang2019midipassage}.  It improves the F measure score from $.843$ to $.889$, while simultaneously reducing the average runtime from $8.01$ seconds down to $0.90$ seconds.

\section{Analysis}
\label{sec:analysis}

In this section, we look at four different questions of interest to gain more intuition about the proposed system.

\subsection{Failure Analysis}
\label{subsec:failureAnalysis}

The first question of interest is: ``Why and where does our system fail?"  To answer this question, we manually inspected all queries where the predicted time interval had no overlap with the ground truth time interval.  These are instances where the system completely failed to find a correct match.  There were four common factors, often co-occuring, that contributed to totally incorrect predictions.  

The first factor is incorrect image re-sizing during the pre-processing stage.  If the initial estimate of staff line separation is incorrect, the re-sized image has a staff line separation that is very different from the expected value.  Because the system has been tuned for a staff line separation of 10 pixels, the system will fail on multiple stages.  One of the reasons for incorrect staff size estimation is that the searchable range of staff sizes was set too narrowly, since the testing set had a wider range of zoom factors than the training set.  This issue could be easily resolved by widening the searchable range, albeit at the cost of increased runtime.  Other reasons for incorrect staff size estimation include substantial blur and images taken at sharp angles (which results in lines that are not horizontal, even locally).  

The second factor is failure from notations that our system does not handle.  These include non-filled noteheads (e.g. half notes, dotted half notes, whole notes), clef changes, 8va octave markings, and trills.  Most of the time, these objects and markings are sufficiently infrequent that our system works well.  In some cases, however, a large frequency of occurrence (e.g. Erik Satie's \textit{Gymnopedies} consists almost entirely of half and dotted half notes) or a combination of these factors (e.g. a clef change in the right hand along with an 8va marking in the left hand) can affect performance.  The clef changes and octave markings could potentially be incorporated into the MIDI bootleg synthesis stage by adding all possible clef and octave interpretations for all note onsets.  We implemented this and ran followup experiments, which are described in the ablation study below (section \ref{subsec:ablation}).  The non-filled noteheads and trills, on the other hand, have no easy, obvious remedy.  Even if we were to replace the simple notehead detector with a state-of-the-art music object detector like the Deep Watershed Detector \cite{tuggener2018deepwatershed}, the authors point out that class imbalance leads to poor detection of infrequent classes such as half notes.

The third factor is stave detection and grouping errors.  One common issue was that the title text for a piece of music (on the first page) would sometimes lead to a lot of spurious detected noteheads, which would in turn lead to a spurious stave in the k-means clustering stage.  The other common issue was staves of music that contained very few detected notes, either because most of the notes were non-filled noteheads or because there were simply very few total notes in the stave.  If the number of detected notes in the stave is sufficiently low, the k-means clustering will fail to form a separate cluster for that stave.

The fourth factor is near-identical matches.  In several instances, the query matched a section of the score that was an almost-perfect match, but differed in perhaps a few measures.  Because of our scoring method, these matches were considered totally incorrect, even though the returned results are very reasonable.  This is an inherent ambiguity in the problem itself, so there is not much that can be done about this.

\begin{table}
	\begin{center}
		\begin{tabular}{| l | c | c |}
			\hline
			System Component & Per Query & Percentage\\
			\hline
			Load MIDI Bootleg Score (III-A) & $.006$s & $\;0.7\%$ \\
			Pre-Processing (III-B1) & $.507$s & $56.6\%$ \\
			Notehead Detection (III-B2) & $.147$s & $16.4\%$ \\
			Staff Line Features (III-B3) & $.088$s & $\;9.8\%$ \\
			Bar Line Features (III-B4) & $.104$s & $11.6\%$ \\
			Query Bootleg Projection (III-B5) & $.028$s & $\;3.2\%$ \\
			Subsequence DTW (III-C) & $.015$s & $\;0.7\%$ \\
			\hline
			Total & $.897$s & $100.0\%$ \\
			\hline
		\end{tabular}
	\end{center}
	\caption{Breakdown of runtime across different system components.  The rightmost columns show the average runtime per query and fraction of total runtime.}
	\label{tab:runtime}
\end{table}

\subsection{Runtime}

The second question of interest is: ``What is the average runtime per query?"  Because the system is an online, user-facing application, the latency is as important of a factor as retrieval accuracy.  The rightmost column of table \ref{tab:results} shows the average runtime per query for the proposed system and baseline systems when run on a $2.1$ GHz Intel Xeon processor.\footnote{The systems based on commercial OMR do not have runtime results because they require the use of a GUI.}  Note that our entire system is implemented in python with OpenCV and a custom cython implementation of subsequence DTW.

There are two things to notice about the runtime results in Table \ref{tab:results}.  First, all of the baseline systems (except for random guessing) are much too slow to be practical for the given application.  This is a consequence of the fact that these baseline systems were originally designed for offline tasks.  Second, the proposed bootleg system is the fastest by far with an average runtime of $0.90$ seconds.  This is more than $200$ times faster than the best performing baseline system (DWD with fine-tuning), while achieving approximately the same retrieval accuracy ($.889$ vs $.891$).  Our proposed system is able to achieve sub-second runtime because we designed our solution from scratch keeping runtime as a primary consideration at each step of the design process.  

Table \ref{tab:runtime} shows the breakdown of total runtime across the different system components described in section \ref{sec:system}.  This profiling analysis reveals that the biggest bottleneck for runtime efficiency is the image pre-processing, which makes up more than $50\%$ of the total runtime.  In fact, about $25\%$ of the total runtime comes from simply decoding the jpg image to raw pixels.  This suggests a straightforward way to further reduce the runtime: when implementing the algorithm on a cell phone, access the raw image pixels before the image is compressed.  By skipping the image compression and decompression, we can reduce the runtime by $25\%$.

\subsection{Feature Size}

The third question of interest is: ``How compact is the bootleg score representation?"  Because the features are transmitted from the cell phone across the network to a server, it is essential for the feature representation to be very compact in order to minimize network latency.  Since the bootleg score is $62$ bixels tall, each column of the bootleg score representation can be encoded as a single 64-bit integer.  Across the 1600 test images, the average size of the bootleg score is 808 bytes, the standard deviation is $304$ bytes, and the maximum size is $2080$ bytes.  Note that these sizes are two orders of magnitude below the initial threshold of 100 KB that we set initially.  The average test image size is $2880 \times 3840$, so the bootleg score representation achieves a compression of more than $40,000\times$ compared to the uncompressed image.

\begin{table}
	\begin{center}
		\begin{tabular}{| l | l | c | l |}
			\hline
			Component & With & Without & Change \\
			\hline
			Adaptive Notehead Estimation & $.889^*$ & $.869$ & $+.020$ \\
			Adaptive Image Resizing  & $.889^*$ & $.785$ & $+.104$ \\
			Background Subtraction & $.889^*$ & $.850$ & $+.039$\\
			Staff Line Re-Estimation & $.889^*$ & $.880$ & $+.009$ \\
			Block Chord Estimation & $.889^*$ & $.860$ & $+.029$ \\
			Filler/Repetition in Bootleg Score & $.889^*$ & $.879$ & $+.010$ \\
			Different Octave Interpretations  & $.890 $ & $\;.889^*$ & $+.001$\\
			Different Clef Interpretations & $.886 $ & $\;.889^*$ & $-.003$ \\
			\hline
		\end{tabular}
	\end{center}
	\caption{Quantifying the importance of various system components.  These results show the test F measure with and without the system component, and the corresponding change in F measure score.  Asterisk indicates the default system described in section \ref{sec:system}.}
	\label{tab:ablation}
\end{table}

\subsection{Ablation Study}
\label{subsec:ablation}

The fourth question of interest is: ``How much do various system components affect retrieval accuracy?"  To answer this question, we conducted an ablation study with several system components.  We performed eight experiments in which we selectively remove a component in order to observe how much it impacts system performance.  The first experiment is to remove the adaptive notehead template estimation.  Instead of adaptively estimating the notehead template on each query image at test time, we instead used a fixed notehead template estimated across all $400$ training images.  The second experiment is to remove the adaptive image resizing.  Instead of estimating the staff size and performing interline normalization, we instead scale the image to a fixed size, regardless of the interline spacing.  The third experiment is to remove the background subtraction, which mitigates the effect of illumination.  The fourth experiment is to remove the staff line re-estimation stage, and only use the initial staff line estimates.  The fifth experiment is to remove the block chord estimation.  In this variant, we only detect isolated floating noteheads that approximately match the notehead template, ignoring larger connected component regions which may correspond to chord blocks containing multiple adjoining notes.  The sixth experiment is to remove filler and repetition columns in the bootleg scores, so that each simultaneous note event only occurs once.  The seventh experiment is to add different octave interpretations of notes in the MIDI-generated bootleg score, in order to correctly handle situations when the sheet music contains an 8va octave marking.  The eighth experiment is to add different clef interpretations of notes in the MIDI-generated bootleg score, in order to correctly handle situations when the sheet music contains clef changes in the left or right hands.  Note that the seventh and eighth experiments \textit{add} rather than remove functionality from the default system described in section \ref{sec:system}.  These two functionalities are explored as a result of the failure analysis in section \ref{subsec:failureAnalysis}.

Table \ref{tab:ablation} shows the results of these eight experiments.  We can group the components into three broad groups: those that have significant effect, those that have moderate effect, and those that have little or no effect.  The adaptive image resizing is the only component with a very large effect, increasing the F measure by $.104$.  Because there is a wide range in the level of zoom among the images, performing interline normalization allows the system to generalize much better.  There are five factors that contribute moderate improvements to system performance: adaptive notehead estimation, background subtraction, staff line re-estimation, block chord estimation, and filler \& repetition in the bootleg score.  The two experimental components (octave \& clef interpretations) had little or no effect on system performance.  It appears that the benefit of handling different octave and clef interpretations is approximately offset by the detriment of spurious matches from adding lots of additional notes to the bootleg score.

\section{Conclusion}
In this article we describe a solution to a challenging multimodal music retrieval problem in which there is a significant runtime constraint and very little training data.  By keeping these two factors as primary considerations throughout the design process, we are able to design a solution that achieves high retrieval accuracy within the given constraints.  With a training set of just 400 cell phone images, we demonstrate that our system generalizes well to a much larger set of 1600 unseen test images, achieves a test F measure score of $0.89$, and has an average runtime of $0.90$ seconds.  Our system is based on a novel feature representation called a bootleg score which encodes the position of noteheads relative to staff lines in sheet music.  This representation encodes the deterministic rules of Western musical notation, and it can be estimated on sheet music images using simple classical computer vision techniques.  The defining characteristic of our system is that it has no trainable weights at all, and its hyperparameters can be tuned easily and intuitively on a small amount of data.  Because of its simplicity and adaptability, we believe that the bootleg score representation may be useful in a variety of multimodal music applications.

There are two areas that we would like to explore in future work.  The first is to relax the assumption that the piece is known.  This would require the system to search a large database of symbolic music files in addition to identifying the corresponding passage in the matching file.  The second area of future work is to explore different applications in which the bootleg score representation might be useful.  This could include camera-based sheet--audio retrieval, sheet music identification, and interactive score applications.

\section*{Acknowledgment}

We gratefully acknowledge the support of NVIDIA Corporation with the donation of the GPU used for this research.

\ifCLASSOPTIONcaptionsoff
  \newpage
\fi



\bibliographystyle{IEEEtran}
\bibliography{IEEEabrv,SheetMidiRetrieval_TMM}
%
%
%

%

\begin{IEEEbiography}[{\includegraphics[width=1in,height=1.25in,clip,keepaspectratio]{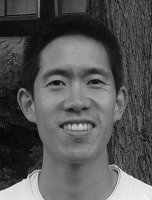}}]{TJ Tsai}
	completed his BS and MS in electrical engineering at Stanford University in 2006 and 2007. From 2008 to 2010, he worked at SoundHound, a startup that allows users to search for music by singing, humming, or playing a recorded track. He completed his Ph.D. at the University of California Berkeley in 2016, and is currently an assistant professor of engineering at Harvey Mudd College.  
\end{IEEEbiography}

\begin{IEEEbiography}[{\includegraphics[width=1in,height=1.25in,clip,keepaspectratio]{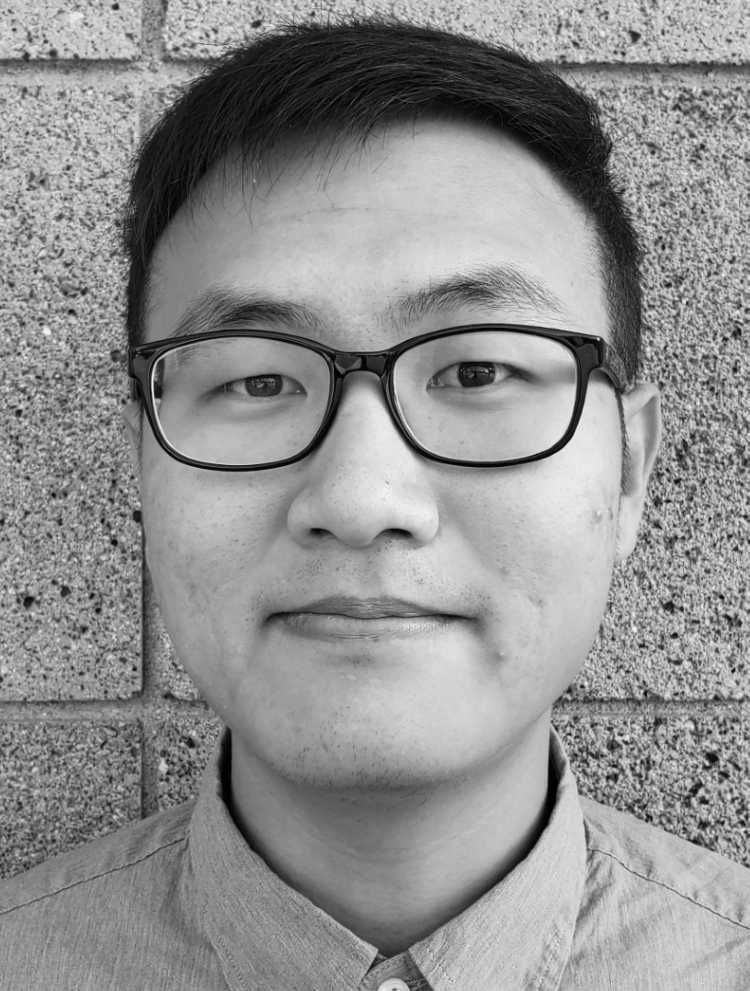}}]{Daniel Yang}
	is currently pursing a BS in computer science and mathematics at Harvey Mudd College. He is interested in computer vision and music information retrieval.
\end{IEEEbiography}

\begin{IEEEbiography}[{\includegraphics[width=1in,height=1.25in,clip,keepaspectratio]{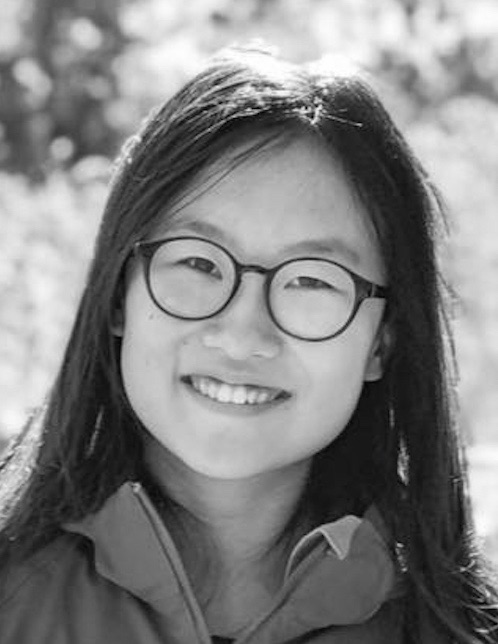}}]{Mengyi Shan}
	is a third-year student at Harvey Mudd College with a double major in mathematics and computer science.
\end{IEEEbiography}

\begin{IEEEbiography}[{\includegraphics[width=1in,height=1.25in,clip,keepaspectratio]{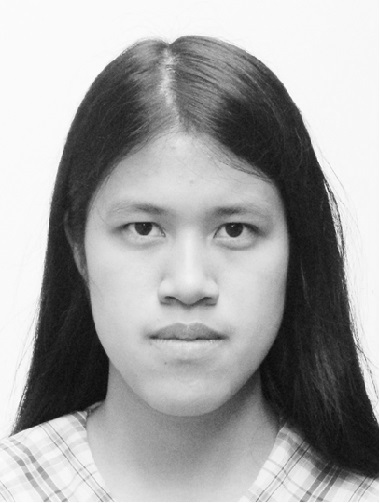}}]{Thitaree Tanprasert}
	is from Bangkok, Thailand. She received a bachelor of science degree from Harvey Mudd College where she joint-majored in computer science and mathematics.  She is currently a graduate student in the Department of Computer Science at University of British Columbia (Vancouver, Canada) and is a member of the Multimodal User eXperience (MUX) lab.
\end{IEEEbiography}

\begin{IEEEbiography}[{\includegraphics[width=1in,height=1.25in,clip,keepaspectratio]{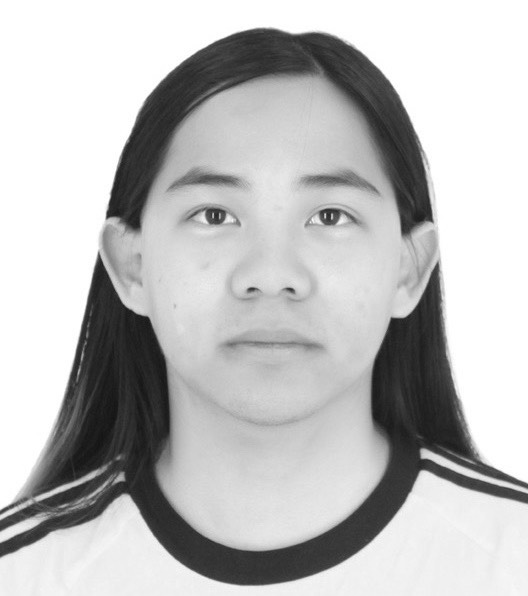}}]{Teerapat Jenrungrot}
	completed his BS in computer science at Harvey Mudd College in 2019. Currently, he is a Ph.D. student at the University of Washington, Seattle in the Department of Computer Science and Engineering, where he is a member of the GRAphics and Imaging Laboratory (GRAIL).
\end{IEEEbiography}








\end{document}